\documentclass[twocolumn,showpacs,aps,prl,superscriptaddress]{revtex4}
\usepackage{graphicx,dcolumn,amsmath,epsfig,color,colordvi,amsfonts} 
\usepackage{fancyhdr,lineno,multirow,rotating,relsize} 
\RequirePackage{xspace}
\def\babar{\mbox{\slshape B\kern-0.1em{\smaller A}\kern-0.1em
    B\kern-0.1em{\smaller A\kern-0.2em R}}}
\def\piz   {\ensuremath{\pi^0}\xspace}
\def\Dbar  {\kern 0.2em\overline{\kern -0.2em D}{}\xspace}
\def\Dz    {\ensuremath{D^0}\xspace}
\def\Dzb   {\ensuremath{\Dbar^0}\xspace}
\def\to    {\ensuremath{\rightarrow}\xspace}
\newcommand{\gevc}{\ensuremath{{\mathrm{\,Ge\kern -0.1em V\!/}c}}\xspace}
\newcommand{\gevcc}{\ensuremath{{\mathrm{\,Ge\kern -0.1em V\!/}c^2}}\xspace}
\newcommand{\mevcc}{\ensuremath{{\mathrm{\,Me\kern -0.1em V\!/}c^2}}\xspace}
\newcommand{\Dzkkpz}{\ensuremath{\Dz \to K^{-}K^{+}\piz}}
\newcommand\fDz{\ensuremath{{f_{\Dz}}}} 

\long\def\inst#1{\par\nobreak\kern 4pt\nobreak
    {\it #1}\par\vskip 10pt plus 3pt minus 3pt}
\def\figurebox#1#2#3{
    \def\arg{#3}
    \ifx\arg\empty
    {\hfill\vbox{\hsize#2\hrule\hbox to 
	#2{\vrule\hfill\vbox to #1{\hsize#2\vfill}\vrule}\hrule}\hfill}
    \else {\hfill\epsfbox{#3}\hfill}
    \fi}
\begin{document}

\begin{flushleft}
\babar-PUB-07/022\\ 
SLAC-PUB-12416\\
arxiv:0704.3593 [hep-ex]
\end{flushleft}

\leftline{Phys. Rev. {\bf D 76} (RC), 011102 (2007)}

\title{{\large \bf \boldmath Amplitude Analysis of the Decay \Dzkkpz}}
%
\author{B.~Aubert}
\author{M.~Bona}
\author{D.~Boutigny}
\author{Y.~Karyotakis}
\author{J.~P.~Lees}
\author{V.~Poireau}
\author{X.~Prudent}
\author{V.~Tisserand}
\author{A.~Zghiche}
\affiliation{Laboratoire de Physique des Particules, IN2P3/CNRS et Universit\'e de Savoie, F-74941 Annecy-Le-Vieux, France }
\author{J.~Garra~Tico}
\author{E.~Grauges}
\affiliation{Universitat de Barcelona, Facultat de Fisica, Departament ECM, E-08028 Barcelona, Spain }
% ...
\collaboration{The \babar\ Collaboration}
\noaffiliation

\date{\today}
\begin{abstract}
Using 385 fb${}^{-1}$ of $e^+e^-$ collisions, 
we study the amplitudes of the singly Cabibbo-suppressed decay \Dzkkpz. We 
measure the strong phase difference between the \Dzb\ and \Dz\ decays to 
$K^*(892)^{+} K^-$ to be $-35.5^\circ\pm1.9^\circ$ (stat) $\pm2.2^\circ$ 
(syst), and their amplitude ratio to be 0.599 $\pm$ 0.013 (stat) $\pm$ 0.011 
(syst). We observe contributions from the $K\pi$ and $K^-K^+$ scalar and 
vector amplitudes, and analyze their angular moments. We find no evidence for 
charged $\kappa$, nor for higher spin states. We also 
perform a partial-wave analysis of the $K^-K^+$ system in a limited mass 
range. 
\end{abstract}
\maketitle
\begin{figure*}[!htbp]
\begin{tabular}{cc} 
\includegraphics[width=0.235\textwidth]{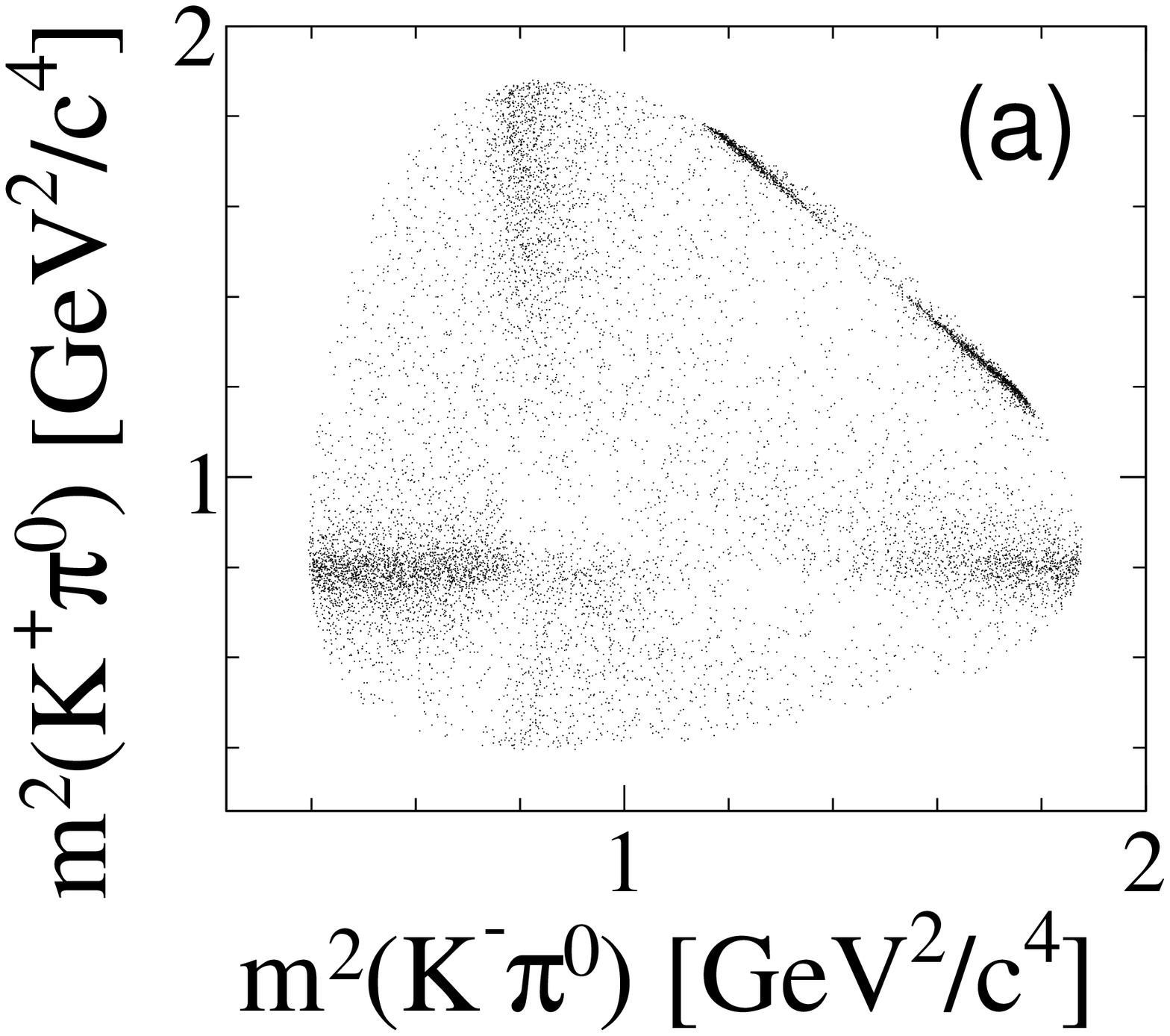}
&
\includegraphics[width=0.235\textwidth]{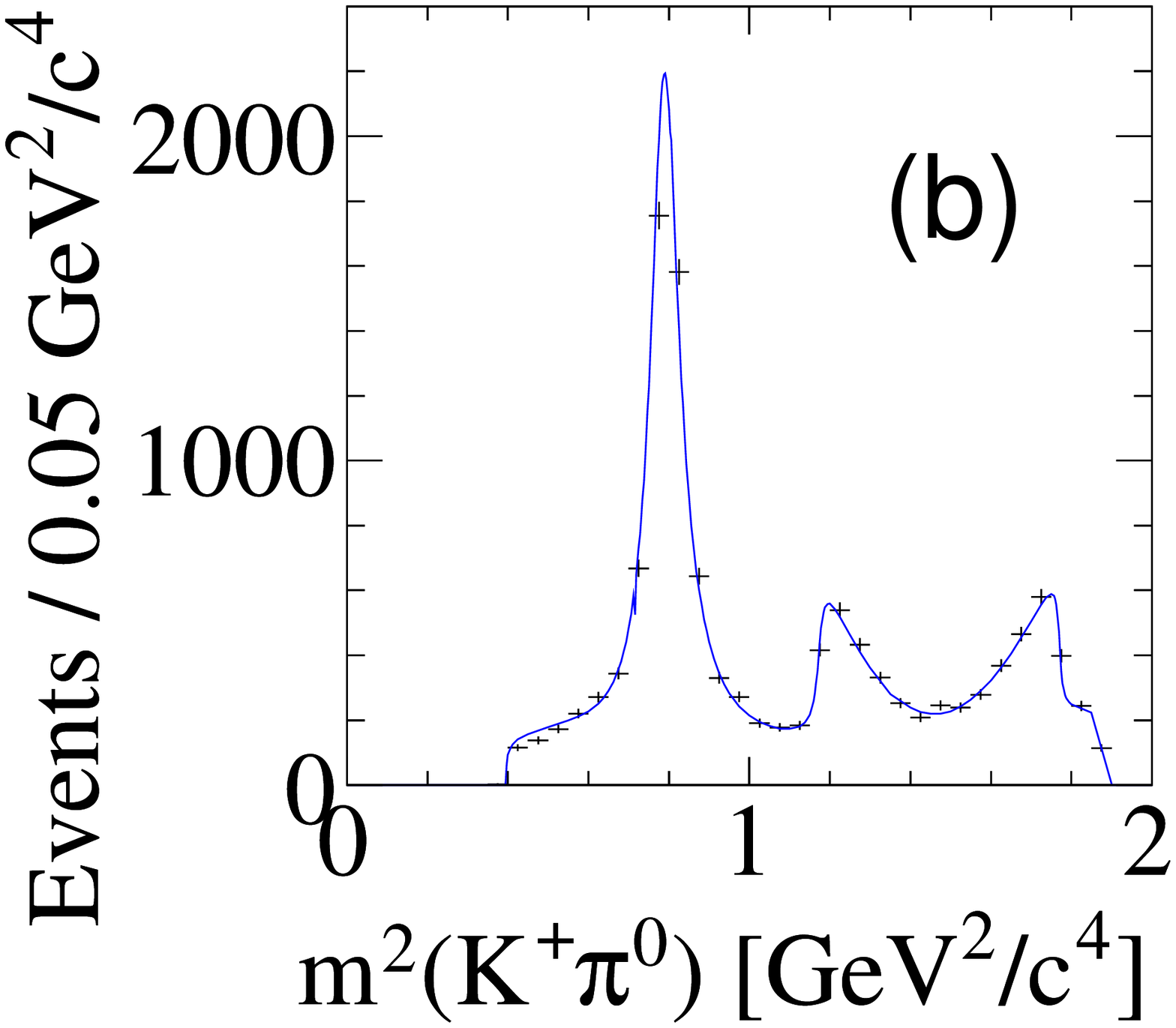}\\
\end{tabular}
\begin{tabular}{cc} 
\includegraphics[width=0.235\textwidth]{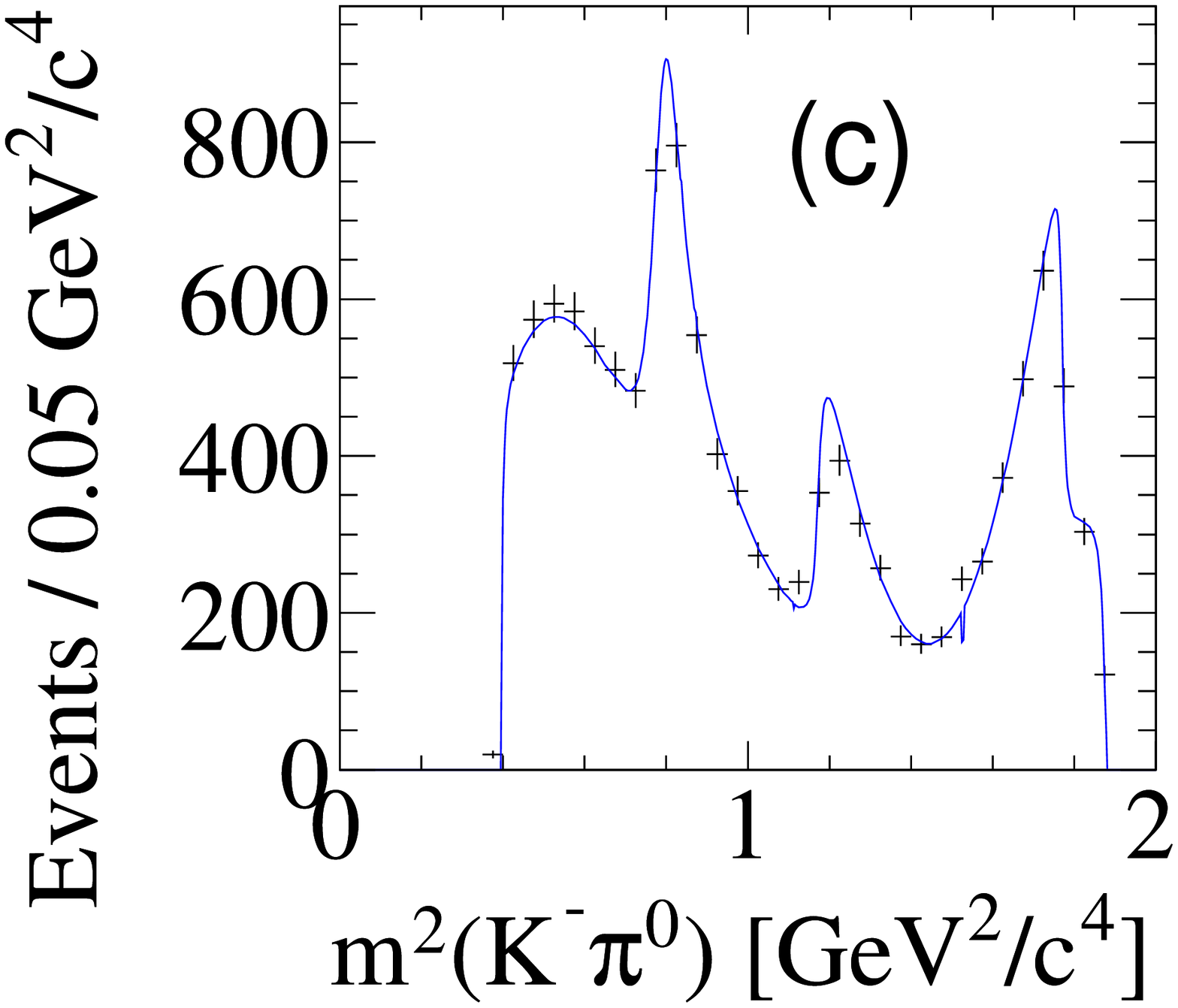}
&
\includegraphics[width=0.235\textwidth]{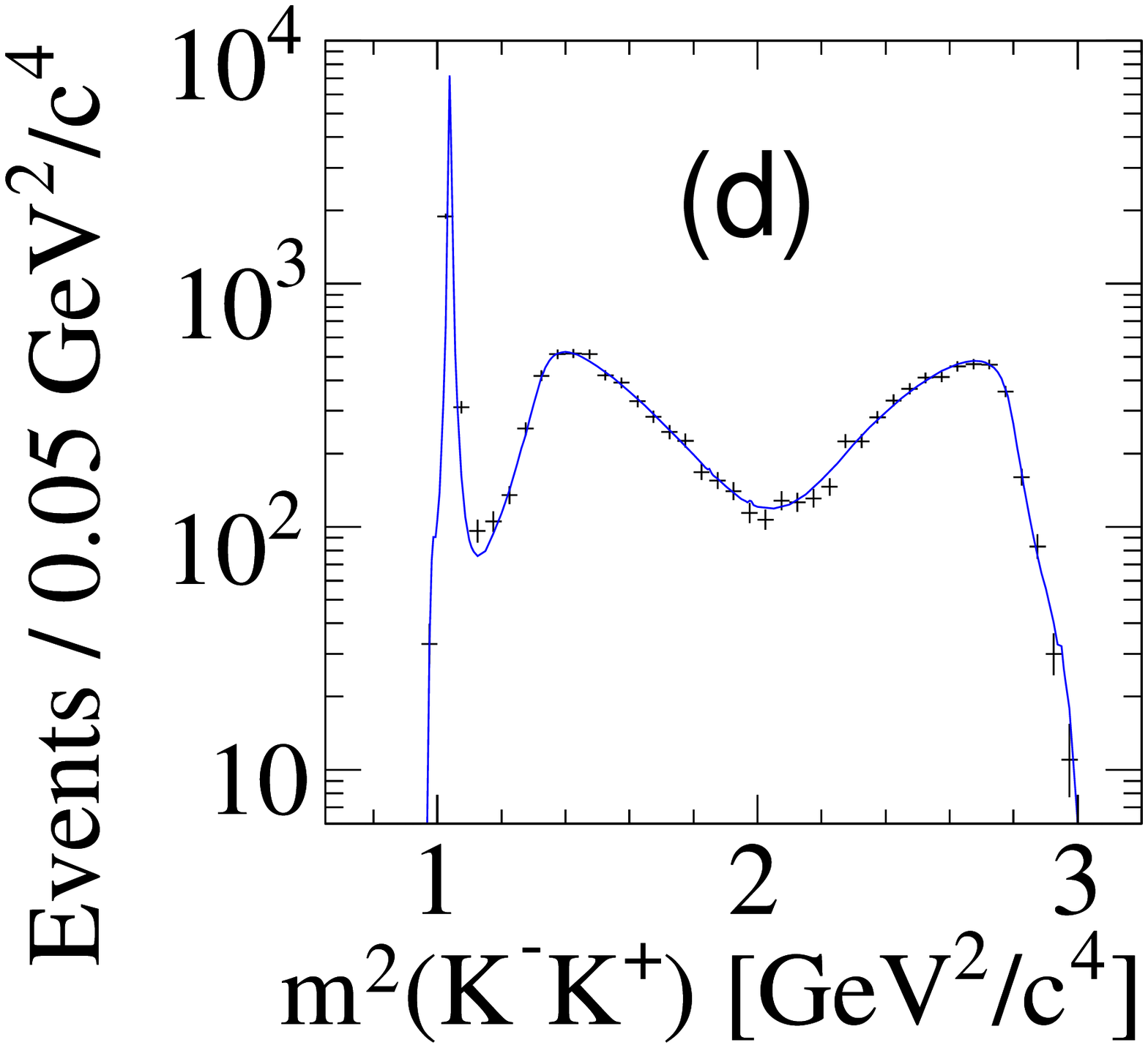}\\
\end{tabular}
\vspace{-1em}
\caption{(Color online) Dalitz plot for $D^0 \to K^- K^+ \piz$~\cite{note1} 
data (a), and the corresponding squared invariant mass projections (b--d). 
The three-body invariant mass of the \Dz\ candidate is constrained to the 
nominal value. In plots~(b--d), the dots (with error bars, black) are data 
points and the solid lines (blue) correspond to the best isobar fit models.}
\label{Fig1}
\end{figure*}
\begin{figure*}[!htbp]
\begin{tabular}{cc} 
\includegraphics[width=0.36\textwidth]{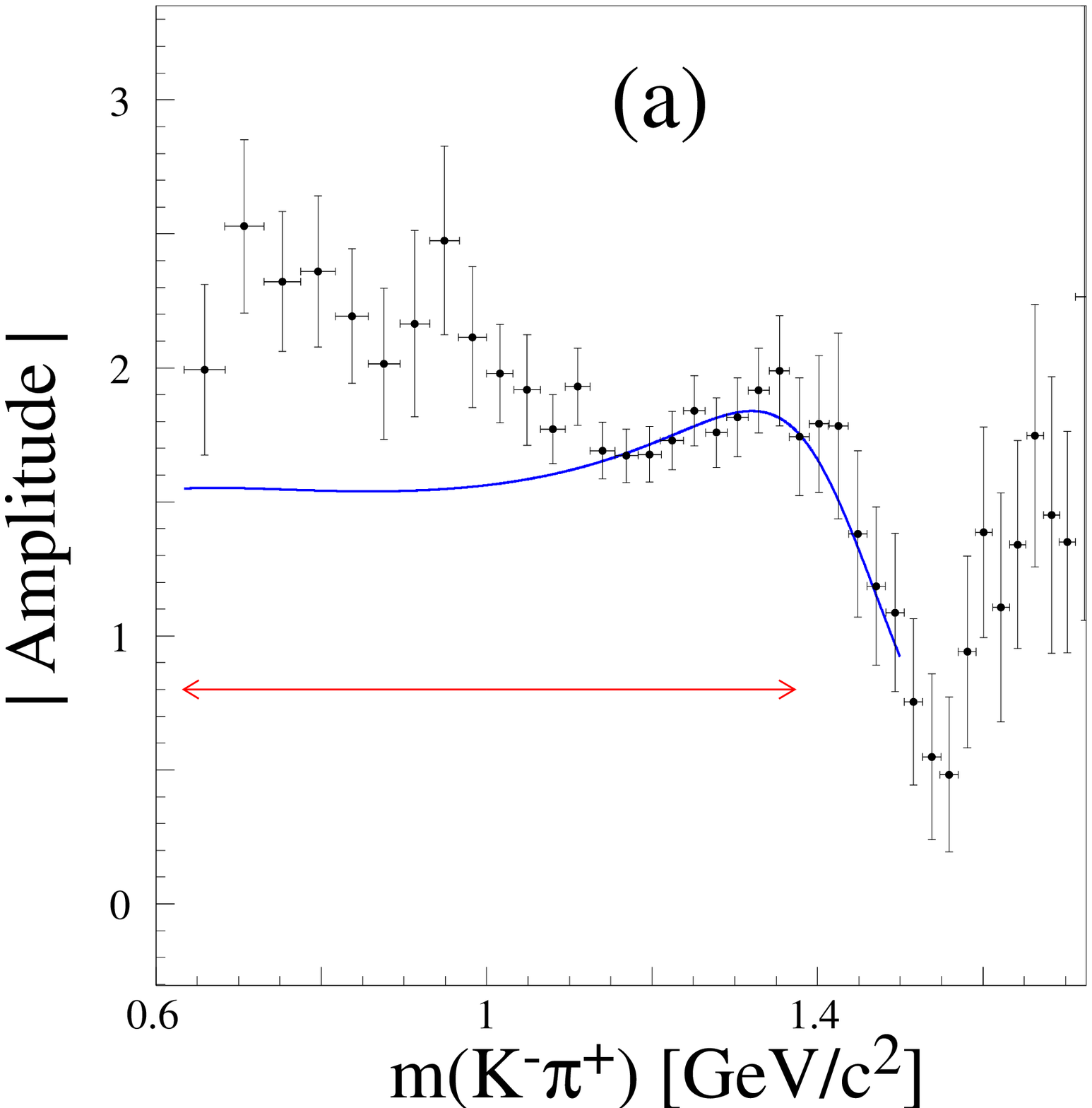}
&
\includegraphics[width=0.36\textwidth]{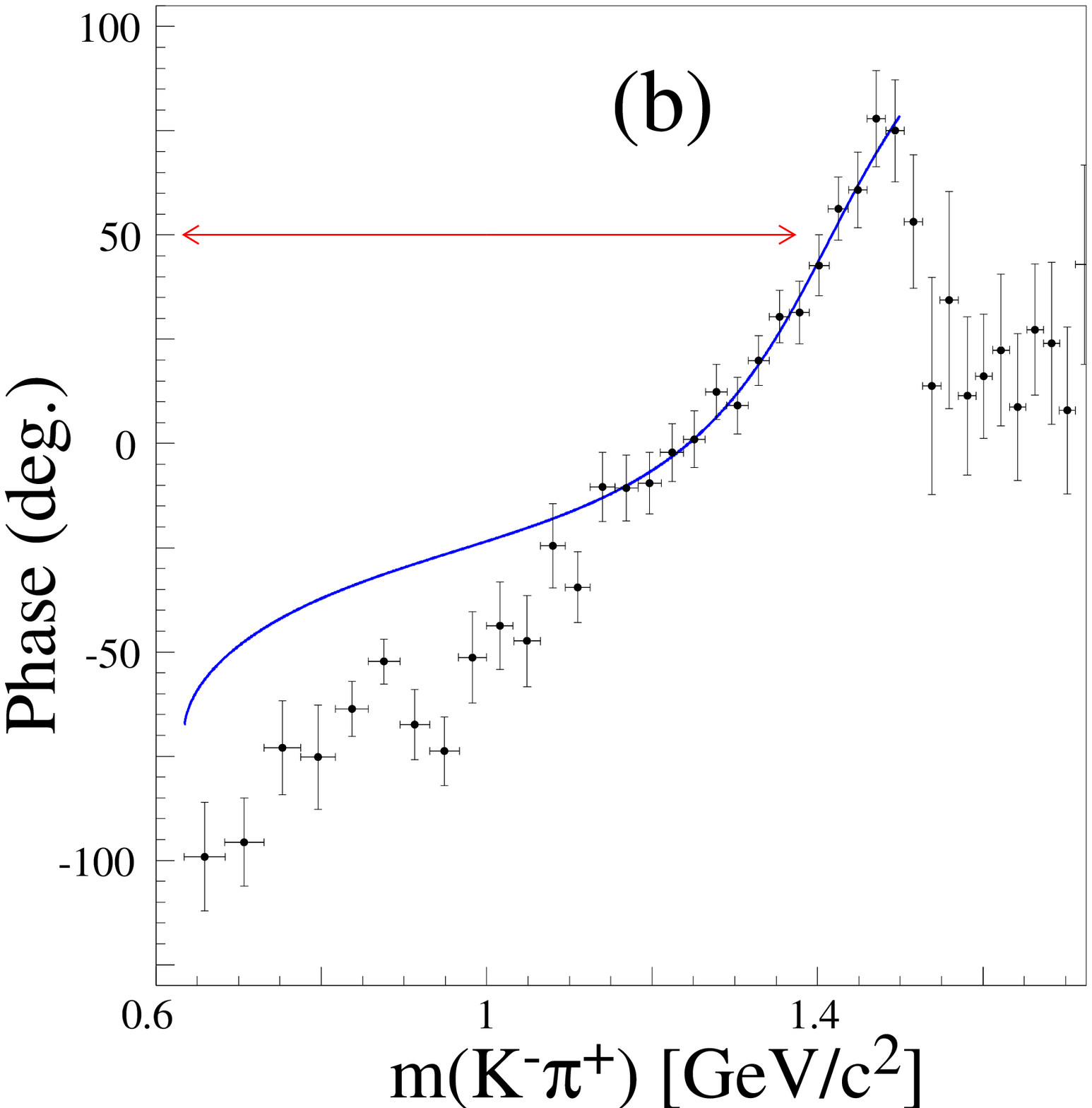} \\
\end{tabular}
\caption{(Color online) LASS (solid line, blue) and E-791 (dots with error 
bars) $K\pi$ \textit{S-}wave amplitudes (a), in arbitrary units, and 
phase (b). The double headed arrow (red) indicates the mass range available 
in the decay \Dzkkpz.}
\label{Fig2}
\end{figure*}
\begin{figure}[!htbp]
\includegraphics[width=3.4in]{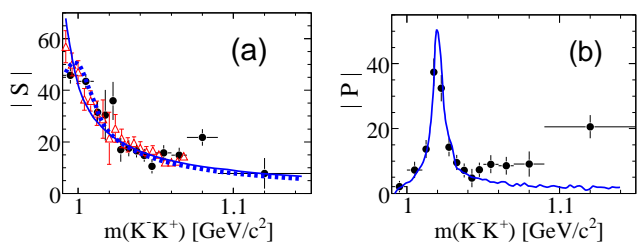}
\vspace{-2em}
\caption{(Color online) The phase-space-corrected $K^-K^+$ \textit{S-} and 
\textit{P-}wave amplitudes, $\left| S \right|$ and $\left| P \right|$ 
respectively, in arbitrary units, as functions of the invariant 
mass. (a) Lineshapes for (solid line, blue) $f_0(980)$, and (broken line, 
blue) $a_0(980)$, derived from Eq.~\ref{eq:7}. (b) Lineshape for $\phi(1020)$ 
(solid line, blue). 
In each plot, solid circles with error bars correspond to values obtained from 
the model-independent analysis for 
$\left| S \right|$ and $\left| P \right|$ using Eq.~\ref{eq:8}.
In (a), the open triangles (red) correspond to values obtained from the decay  
$\Dz\to K^-K^+\bar{K^0}$ (see text).} 
\label{Fig3}
\end{figure}

The amplitudes describing $D$ meson weak decays into three-body final states 
are dominated by intermediate resonances that lead to highly nonuniform 
intensity distributions in the available phase space. Analyses of these 
distributions have led to new insights into the role of the light-meson 
systems produced~\cite{motivation}. The $K^{\pm}\piz$ systems from the decay 
\Dzkkpz~\cite{note1} can provide information on the $K\pi$ \textit{S-}wave 
(spin-0) amplitude in the mass range 0.6--1.4 \gevcc, and hence on the 
possible existence of the $\kappa(800)$, reported to date only in the neutral 
state ($\kappa^0 \to K^- \pi^+$)~\cite{kappa}. If the $\kappa$ has isospin 
$1/2$, it should be observable also in the charged states. Results of the 
present analysis can be an input for extracting the $C\!P$-violating phase  
$\gamma = \arg{\left(- V^{}_{ud} V_{ub}^\ast/ V^{}_{cd} V_{cb}^\ast\right)}$ 
of the quark mixing matrix by exploiting interference structure in the Dalitz 
plot from the decay $B^{\pm}\to\Dz_{K^-K^+\piz}K^{\pm}$~\cite{abi, myGamma}. 
Singly Cabibbo-suppressed decays are also important because they might be 
sensitive to direct $C\!P$ violation in charm decays~\cite{kagan}, the 
discovery of which might indicate physics beyond the Standard Model.\\
\indent We perform the present analysis on 385 fb${}^{-1}$ of $e^+e^-$ 
collision data collected at and around 10.58~GeV center-of-mass (CM) energy
with the \babar\ detector~\cite{detector} at the PEP-II storage ring. 
We distinguish \Dz from \Dzb by reconstructing  
the decays $D^{*+}\to\Dz\pi^{+}$ and $D^{*-}\to\Dzb\pi^{-}$. The 
event-selection criteria are the same as those used in our measurement of 
the branching ratio of the decay \Dzkkpz~\cite{mybr}. In particular, we 
require that the CM momentum of $D^0$ candidate be greater than 
2.77~\gevc, and that $|m_{D^{*+}} - m_{D^0} - 145.4| < 0.6~\mevcc$, where 
$m$ refers to a reconstructed invariant mass. To minimize uncertainty 
from background shape, we choose a sample of very high purity ($\sim 98.1\%$) 
using $1855 < m_{D^0} < 1875$~\mevcc, and find $11278 \pm 110$ signal events.  
We estimate the signal efficiency for each event as a function of its 
position in the Dalitz plot using simulated \Dzkkpz\ events from 
$c\overline c$ decays, generated uniformly in the available phase space. To 
correct for differences in particle-identification rates  
in data and simulation, we determine the ratio of these for each track, and 
apply an event-by-event correction factor.\\
\indent Neglecting $C\!P$ violation in $D$ meson decays, we define the 
\Dz (\Dzb) decay amplitude $\cal{A}$ ($\bar{\cal{A}}$) 
in the \Dzkkpz\ Dalitz plot of Fig.~\ref{Fig1}, as:
\begin{eqnarray}
\label{eq:1}
   {\cal{A}}[\Dz\to K^- K^+\piz] \equiv \fDz(m_{K^+\piz}^2, m_{K^-\piz}^2),\\
\label{eq:2}
 \bar{\cal{A}}[\Dzb\to K^+ K^-\piz] \equiv \fDz(m_{K^-\piz}^2, m_{K^+\piz}^2).
\end{eqnarray}
The complex quantum mechanical amplitude $f$ is a coherent sum of all relevant 
quasi-two-body $\Dz\to(r\to AB)C$ isobar model~\cite{isobar} resonances, 
$f = \sum_r a_r e^{i\phi_r} A_r(s)$. Here $s=m_{AB}^2$, and $A_r$ is the 
resonance amplitude. We obtain coefficients $a_r$ and $\phi_r$ from 
a likelihood fit. The probability density function for signal events 
is $\left| f \right|^2$. We model incoherent background empirically using 
events from the lower sideband of the $m_{D^0}$~\cite{mybr} distribution.\\
\indent For \Dz\ decays to spin-1 (\textit{P-}wave) and spin-2 states, we 
use the Breit-Wigner amplitude,
\begin{eqnarray}
\label{eq:3}
A_{BW}(s) &=& {{\cal{M}}_L(s,p)}\hspace{2pt}{1\over{M_0^2-s-iM_0\Gamma(s)}},\\
\label{eq:4}
\Gamma(s) &=& \Gamma_0\Big({M_{0}\over \sqrt{s}}
\Big)\Big({p\over{p_{0}}}\Big)^{2L+1} {\Big[{{\cal{F}}_L(p) \over 
      {\cal{F}}_L(p_0)}\Big]^2},
\end{eqnarray}
\noindent where $M_0$ ($\Gamma_0$) is the resonance mass (width)~\cite{pdg}, 
$L$ is the angular momentum quantum number, $p$ is the momentum of either 
daughter in the resonance rest frame, and $p_0$ is the value of $p$ when s = 
$M_0^2$. The function ${\cal{F}}_L$ is the  Blatt-Weisskopf barrier 
factor~\cite{bw}: ${{{\cal{F}}_0}}$ = 1, ${{{\cal{F}}_1}}$ = 
$1/\sqrt{ 1+ {Rp}^2}$, and ${{{\cal{F}}_2}}$ = 
$1/\sqrt{ 9+ 3 {Rp}^2 + {Rp}^4}$, where we take the meson radial parameter $R$ 
to be 1.5 GeV${}^{-1}$~\cite{valR}. We define the spin part of the amplitude, 
${{\cal{M}}_L}$, as: ${{\cal{M}}_0}$ = $M_{\Dz}^2$, ${{\cal{M}}_1}$ = 
-2 $\vec{p_A}.\vec{p_C}$, and ${{\cal{M}}_2}$ = $4\over 3$ [
$3{(\vec{p_A}.\vec{p_C})}^2 - {|\vec{p_A}|}^2.{|\vec{p_C}|}^2$] 
$M_{\Dz}^{-2}$, where $M_{\Dz}$ is the nominal \Dz\ mass, and $\vec{p_i}$ 
is the 3-momentum of particle $i$ in the resonance rest frame.\\  
\indent For \Dz\ decays to $K^{\pm}\piz$ \textit{S-}wave states, we consider three 
amplitude models. One model uses the LASS amplitude for  
$K^-\pi^+\to K^-\pi^+$ elastic scattering~\cite{LASS},
\begin{eqnarray}
  \label{eq:5}
  A_{{K\pi}(S)} (s) = {\sqrt{s} \over p}\sin\delta(s) e^{i\delta(s)},\indent{}
\indent{}\indent{}\indent{}\indent{}\indent{}\indent{}\indent{}\indent{}\\
  \label{eq:6}
  \delta(s) = \cot^{-1}\Big({1\over pa} + {bp\over 2}\Big) 
              + \cot^{-1}\Big({M_{0}^2-s\over M_{0}\Gamma_{0}\cdot{M_{0}\over 
		  \sqrt{s}}\cdot{p\over{p_{0}}}}\Big),
\end{eqnarray}
\noindent where $M_{0}$ ($\Gamma_{0}$) refers to the $K^*_0(1430)$ mass 
(width), $a =1.95 \pm 0.09$ GeV${}^{-1}c$, and $b = 1.76 \pm 0.36$ 
GeV${}^{-1}c$. The unitary nature of Eq.~\ref{eq:5} provides a good 
description of the amplitude up to 1.45 \gevcc\ (i.e., $K\eta^{\prime}$ 
threshold). In Eq.~\ref{eq:6}, the first term is a nonresonant contribution 
defined by a scattering length $a$ and an effective range $b$, and the 
second term represents the $K^*_0(1430)$ resonance. The phase space factor 
$\sqrt{s}/ p$ converts the scattering amplitude to the invariant amplitude. 
Our second model uses the E-791 results for the $K^-\pi^+$ \textit{S-}wave 
amplitude from an energy-independent partial-wave analysis in the 
decay $D^+\to K^-\pi^+\pi^+$~\cite{brian}. The third model uses a coherent 
sum of a uniform nonresonant term, and Breit-Wigner terms for the 
$\kappa(800)$ and $K^*_0(1430)$ resonances.\\
\indent In Fig.~\ref{Fig2} we compare the $K\pi$ \textit{S-}wave amplitude 
from the E-791 analysis~\cite{brian} to the LASS amplitude of 
Eqs.~\ref{eq:5}--\ref{eq:6}. For easy comparison, we have normalized the 
LASS amplitude in Fig.~\ref{Fig2}a approximately to the E-791 measurements 
with $\sqrt{s} > 1.15 \gevcc$, and have reduced the LASS phase, $\delta(s)$, 
in Fig.~\ref{Fig2}b by $80^\circ$. We then observe good agreements in the 
mass dependence of amplitude and phase for $\sqrt{s} > 1.15 \gevcc$. As the 
mass decreases from 1.15~\gevcc, the E-791 amplitude increases while the LASS 
amplitude decreases, with the ratio finally reaching $\sim$1.7 at threshold. 
At the same time, their phase difference increases to $\sim$$40^\circ$ at 
threshold. This behavior might be due to the form factor describing \Dz 
decay to a $K\pi$ \textit{S-}wave system and a bachelor $\bar{K}$. Since no 
centrifugal barrier is involved, such an effect should be more significant 
for \textit{S-}wave than for higher spin waves because of the larger overlap 
between the initial and final state wave functions. However, the inverse 
momentum of the $K\pi$ system in the \Dz rest frame increases 
from 0.27 Fermi at $K\pi$ threshold to 0.48 Fermi at 1.15 \gevcc, therefore 
any form factor effect would decrease with increasing $K\pi$ mass. If the 
effect is essentially gone by 1.15 \gevcc, similar mass dependence of 
amplitude and phase in \Dz\ decay and $K\pi$ scattering would be observable 
at higher mass values, in agreement with Fig.~\ref{Fig2}. In the present 
analysis, we make an attempt to distinguish between the two rather different 
$K\pi$ \textit{S-}wave mass dependences in the region below $\sim$1.15 \gevcc. 
In each case, we also allow the fit to determine the strength and phase of 
these amplitudes relative to the $K^*(892)^{+}$ reference.\\
\indent We describe the \Dz decay to a $K^-K^+$ \textit{S-}wave state by a 
coupled-channel Breit-Wigner amplitude for the $f_0(980)$ and $a_0(980)$ 
resonances, with their respective couplings to $\pi\pi$, $K \bar{K}$ and 
$\eta\pi$, $K\bar{K}$ final states~\cite{flatte},
\begin{equation}
A_{f_0[a_0]} (s) = \frac{M_{\Dz}^2} {M_{0}^2 - s - 
  i (g_1^2~\rho_{\pi\pi\left[\eta\pi\right]} + g_2^2~\rho_{K\bar{K}})}.
\label{eq:7}
\end{equation}
\noindent Here $\rho$ represents Lorentz invariant phase space, $2p/\sqrt{s}$. 
For the $f_0(980)$, we use the BES~\cite{bes} parameter values 
$M_{0} = $ 965$\pm$10~\mevcc, $g_1^2 = $ 165$\pm$18~MeV${}^{2}$/c${}^{4}$, and 
${g_2^2}/{g_1^2} = $ 4.21$\pm$0.33. For the $a_0(980)$, we use the Crystal 
Barrel~\cite{crystalbarrel} values $M_{0} = $ 999$\pm$2~\mevcc,  
$g_1 = $ 324$\pm$15~\mevcc, and ${g_1^2}/{g_2^2} = $ 1.03$\pm$0.14.
Only the high mass tails of $f_0(980)$ and $a_0(980)$ are observable, as 
shown in Fig.~\ref{Fig3}a.  They are similar, so we try a model for each as a 
description of the $K^-K^+$ \textit{S-}wave amplitude. In Fig.~\ref{Fig3}b we 
show, in the same mass range, the $K^-K^+$ \textit{P-}wave amplitude 
parametrized by the $\phi(1020)$ meson.\\
\begin{table*}[htbp]
\caption{The results obtained from the $D^0 \to K^- K^+ \piz$ Dalitz plot 
fit. We define amplitude coefficients, $a_r$ and $\phi_r$, relative to those 
of the $K^{*}(892)^+$. The errors are statistical and systematic, 
respectively. We show the $a_0(980)$ contribution, when it is included in 
place of the $f_0(980)$, in square brackets. We denote the 
$K\pi$ \textit{S-}wave states here by $K^{\pm}\piz(S)$. We use LASS amplitude 
to describe the $K\pi$ \textit{S-}wave states in both the isobar models 
(I and II).}
\label{tab:result}
\begin{tabular}{|l|rrr|rrr|}
\hline
          &&   Model I &   &&  Model II & \cr
\hline
State &   Amplitude, $a_r$ &  Phase, $\phi_r$ (${}^\circ$) & Fraction, 
 $f_r$ (\%) &   Amplitude, $a_r$ &  Phase, $\phi_r$ (${}^\circ$) &  
 Fraction, $f_r$ (\%)\cr 
 \hline\hline
$K^*(892)^{+}$ & 1.0 (fixed) & 0.0 (fixed) & 45.2$\pm$0.8$\pm$0.6 & 1.0 (fixed)
& 0.0 (fixed) &  44.4$\pm$0.8$\pm$0.6\cr
$K^*(1410)^{+}$ &  2.29$\pm$0.37$\pm$0.20  &   86.7$\pm$12.0$\pm$9.6  
&  3.7$\pm$1.1$\pm$1.1 &   &    & \cr
$K^+\piz(\textit{S})$  &  1.76$\pm$0.36$\pm$0.18 & -179.8$\pm$21.3$\pm$12.3 
&  16.3$\pm$3.4$\pm$2.1&  3.66$\pm$0.11$\pm$0.09  & -148.0$\pm$2.0$\pm$2.8   
&  71.1$\pm$3.7$\pm$1.9\cr
$\phi(1020)$ &  0.69$\pm$0.01$\pm$0.02  &  -20.7$\pm$13.6$\pm$9.3  
&  19.3$\pm$0.6$\pm$0.4 &  0.70$\pm$0.01$\pm$0.02  
&   18.0$\pm$3.7$\pm$3.6   &  19.4$\pm$0.6$\pm$0.5\cr
$f_0(980)$ &  0.51$\pm$0.07$\pm$0.04  & -177.5$\pm$13.7$\pm$8.6  
&  6.7$\pm$1.4$\pm$1.2 &  0.64$\pm$0.04$\pm$0.03  &  -60.8$\pm$2.5$\pm$3.0 
&  10.5$\pm$1.1$\pm$1.2\cr
$\left[a_0(980)^0\right]$ & [0.48$\pm$0.08$\pm$0.04]& [-154.0$\pm$14.1$\pm$8.6]
&  [6.0$\pm$1.8$\pm$1.2] &  [0.68$\pm$0.06$\pm$0.03]& [-38.5$\pm$4.3$\pm$3.0] 
&  [11.0$\pm$1.5$\pm$1.2]\cr
$f_2'(1525)$  &  1.11$\pm$0.38$\pm$0.28  &  -18.7$\pm$19.3$\pm$13.6 &  
0.08$\pm$0.04$\pm$0.05 &   &  &  \cr
$K^*(892)^{-}$ &  0.601$\pm$0.011$\pm$0.011 &  -37.0$\pm$1.9$\pm$2.2   
&  16.0$\pm$0.8$\pm$0.6 &  0.597$\pm$0.013$\pm$0.009 
&  -34.1$\pm$1.9$\pm$2.2   &  15.9$\pm$0.7$\pm$0.6\cr
$K^*(1410)^{-}$ &  2.63$\pm$0.51$\pm$0.47  & -172.0$\pm$6.6$\pm$6.2   
&  4.8$\pm$1.8$\pm$1.2 &   &    & \cr
$K^-\piz(\textit{S})$& 0.70$\pm$0.27$\pm$0.24 & 133.2$\pm$22.5$\pm$25.2 
& 2.7$\pm$1.4$\pm$0.8 &  0.85$\pm$0.09$\pm$0.11  &  108.4$\pm$7.8$\pm$8.9 
& 3.9$\pm$0.9$\pm$1.0\cr
\hline 
\end{tabular}
\end{table*}
\indent To fit the Dalitz plot, we try several models incorporating  
various combinations of intermediate states. 
In each fit, we include the $K^*(892)^{+}$ and measure the complex amplitude 
coefficients of other states relative to it. 
As a check on the quality of each fit, we compare the number of 
events observed in bins in the Dalitz plot with the number predicted 
by the fit.  We compute residuals and statistical uncertainties 
to form a $\chi^2$, and take $\chi^2/\nu$ (where $\nu$ is the number 
of bins less number of variable parameters) as a figure of merit. We also 
compare the distributions of angular moments (described later) predicted 
by the fit and actually observed in the data.\\
\indent The LASS $K\pi$ \textit{S-}wave amplitude gives the best agreement 
with data and we use it in our nominal fits (see next paragraph). 
The $K\pi$ \textit{S-}wave modeled by the combination of $\kappa(800)$ (with 
parameters taken from Ref.~\cite{kappa}), a nonresonant term and 
$K^*_0(1430)$ has a smaller fit probability ($\chi^2$ probability $<$ 5\%). 
The best fit with this model ($\chi^2$ probability 13\%) yields a charged 
$\kappa$ of mass (870 $\pm$ 30)~\mevcc, and width (150 $\pm$ 20)~\mevcc, 
significantly different from those reported in Ref.~\cite{kappa} for the 
neutral state. This does not support the hypothesis that production of a 
charged, scalar $\kappa$ is being observed. The E-791 amplitude~\cite{brian} 
describes the data well, except near threshold ($\chi^2$ probability  23\%). 
Though our data favor the LASS parametrization for $\sqrt{s} < 1.15 \gevcc$, 
the insensitivity of the fit to small variations in amplitude at these masses 
does not allow an independent \textit{S-}wave measurement with the present 
data sample. Therefore, we use the E-791 amplitude to estimate systematic 
uncertainty in our results.\\ 
\begin{figure*}[!htbp]
\begin{tabular}{cc} 
\includegraphics[width=0.505\textwidth]{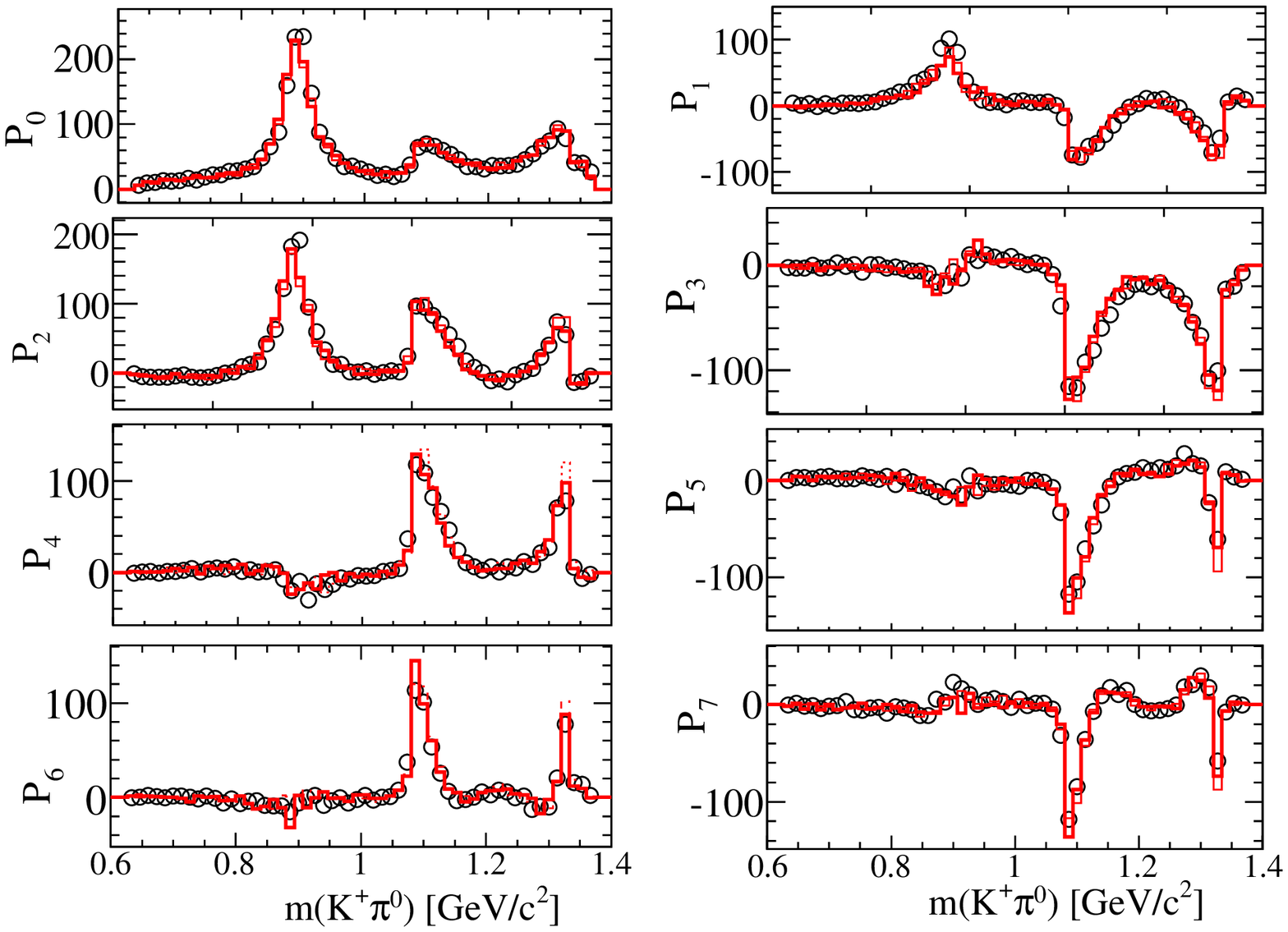} &
\includegraphics[width=0.505\textwidth]{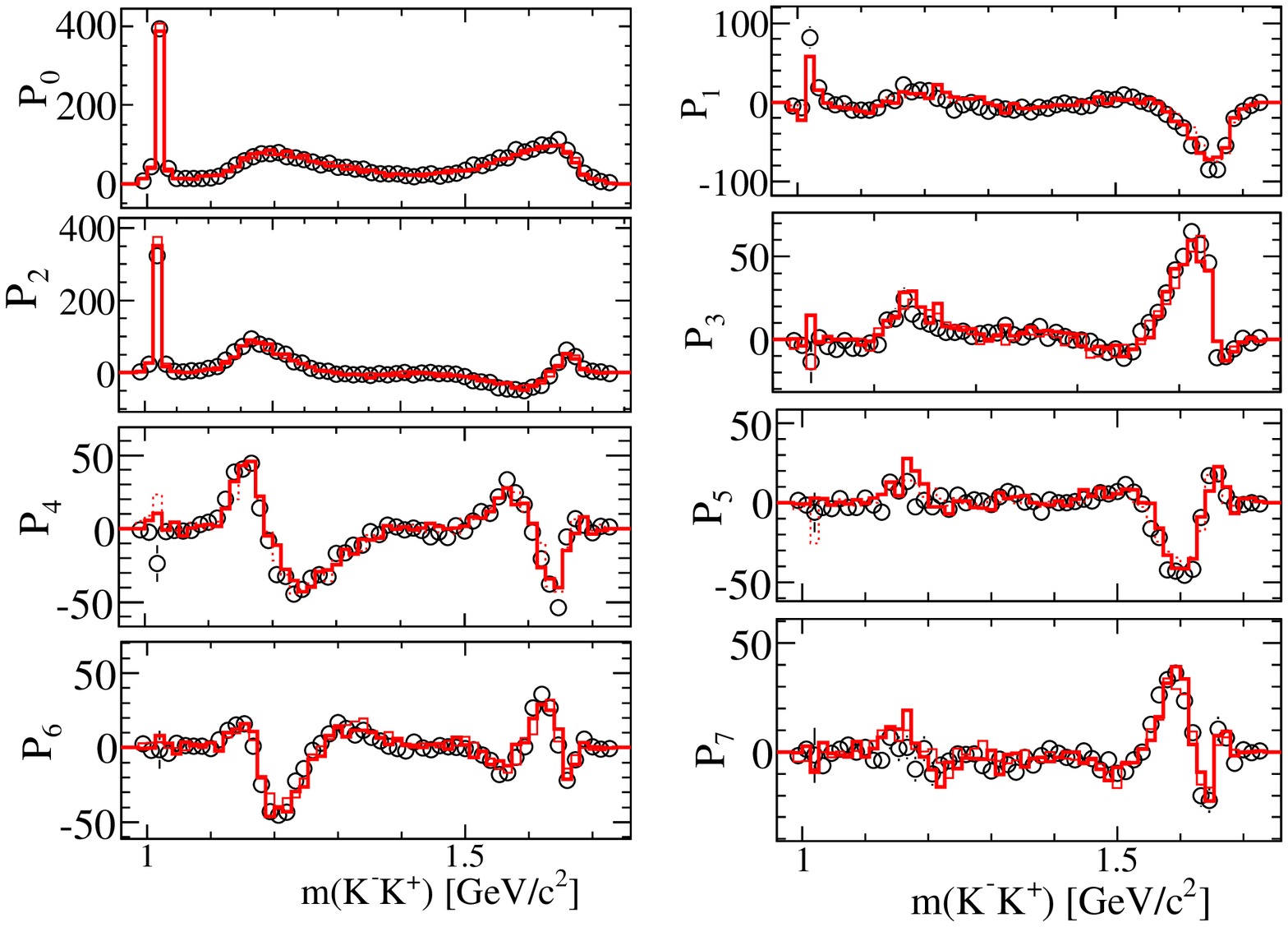} \\
\end{tabular}
\caption{(Color online) The mass dependence of the spherical harmonic moments 
of $\cos\theta_H$ after efficiency corrections and background subtraction: 
$K^+\piz$ (columns I, II) and $K^-K^+$ (columns III, IV). The circles with 
error bars are data points and the curves (red) are derived from the fit 
functions (see text). For the sake of visibility, we do not show error bars on 
the curves.} 
\label{Fig4}
\end{figure*}
\indent We find that two different isobar models describe the data well. Both 
yield almost identical behavior in invariant mass 
(Fig.~\ref{Fig1}b--\ref{Fig1}d) and angular distribution (Fig.~\ref{Fig4}). 
We use LASS amplitude to describe the $K\pi$ \textit{S-}wave amplitudes in 
both the isobar models (I and II).
We summarize the results of the best fits (Model I: $\chi^2/\nu = 702.08/714$, 
probability 61.9\%; Model II:  $\chi^2/\nu = 718.89/717$, probability 47.3\%)
in Table~\ref{tab:result}. We also list the fit fraction for each resonant 
process $r$, defined as 
{\footnotesize{ $f_r \equiv \int |a_r A_r|^2 d\tau / \int |\fDz|^2 d\tau$,}}
where $d\tau = d m_{K^-\piz}^2 d m_{K^+\piz}^2$, in Table~\ref{tab:result}. 
Due to interference among the contributing amplitudes, the $f_r$ do 
not sum to one in general. We find that the $K\pi$ \textit{S-}wave is not in 
phase with the \textit{P-}wave at threshold as it was in the LASS scattering 
data. For Model I (II), the \textit{S-}wave phase relative to the 
$K^*(892)^{+}$ is $\sim$$180^\circ$ $(150^\circ)$ for the positive charge and 
$135^\circ$ $(110^\circ)$ for the negative charge.\\ 
\indent We have also considered the possible contributions from other resonant 
states such as: $K_2^*(1430)$, $f_2(1270)$, $f_0(1370)$, and $f_0(1510)$. We 
find that none of them is needed to describe the Dalitz plot, they all provide 
small contributions and lead to smaller $\chi^2$ probabilities.\\
\indent Angular distributions provide a more detailed information on specific 
features of the amplitudes used in the description of the Dalitz plot. We 
define the helicity angle $\theta_H$ for the decay $\Dz\to(r\to AB)C$ as the 
angle between the momentum of $A$ in the $AB$ rest frame and the momentum of 
$AB$ in the $D^0$ rest frame. The moments of $\cos\theta_H$, defined as the 
efficiency-corrected and background-subtracted invariant mass distributions 
of events weighted by spherical harmonic functions, 
$Y_l^0(\cos\theta_H) = \sqrt{{2l+1} \over {4\pi}}~ P_l(\cos\theta_H)$, 
where the $P_l$ are Legendre polynomials of order $l$, are shown in 
Fig.~\ref{Fig4} for the $K^+ \piz$ and $K^- K^+$ channels, for $l = 0-7$. The 
$K^- \piz$ moments are similar to those for $K^+ \piz$.\\ 
\indent The mass dependent $K^-K^+$ \textit{S-} and \textit{P-}wave complex 
amplitudes can also be obtained directly from our data in a model-independent 
way in a limited mass range around 1~\gevcc. In a region of the Dalitz plot 
where \textit{S-} and \textit{P-}waves in a single channel dominate, their 
amplitudes are given by the following Legendre polynomial moments,
\begin{equation}
\label{eq:8}
\footnotesize{ 
P_0 = \frac{{\left| S \right|}^2 + {\left| P \right|}^2}{\sqrt{2}},~ 
P_1 = {\sqrt{2}}{\left| S \right|}{\left| P \right|}~\cos\theta_{SP},~
P_2 = {\sqrt{2 \over5}} ~\left| P \right|^2,
}
\end{equation}
\noindent using $\int\limits_{-1}^1{P_l P_m d(\cos{\theta_H})} = \delta_{lm}$. 
Here $\left| S \right|$ and $\left| P \right|$ are, respectively, the 
magnitudes of the \textit{S-} and \textit{P-}wave amplitudes, and 
$\theta_{SP} = \theta_\textit{S} - \theta_P$ is the relative phase between 
them. We use these relations to evaluate 
$\left| S \right|$  and $\left| P \right|$, shown in Fig.~\ref{Fig3}, for the 
$K^-K^+$ channel in the mass range $m_{K^-K^+} < 1.15~\gevcc$. The measured 
values of $\left| S \right|$ agree well with those obtained in the analysis 
of the decay $\Dz\to K^-K^+\bar{K^0}$~\cite{antimo}. They also agree well with 
either the $f_0(980)$ or the $a_0(980)$ lineshape. The measured values of 
$\left| P \right|$ are consistent with a Breit-Wigner lineshape for 
$\phi(1020)$. Results for $\cos\theta_{SP}$ and $\theta_{SP}$ are shown in 
Figs.~\ref{Fig5}a--\ref{Fig5}b. A twofold ambiguity in the 
sign of $\theta_{SP}$ exists, as shown in Fig.~\ref{Fig5}b. It is, however, 
straightforward to choose the physical solution. In this region, the 
$\phi(1020)$ meson (\textit{P-}wave) has a very rapidly rising phase, while 
we expect the \textit{S-}wave phase to be relatively slowly varying. Thus, the 
upper solution, in which $\theta_\textit{S}-\theta_P$ is rapidly falling, is 
the physical solution. We take the Breit-Wigner phase of $\phi(1020)$, shown 
in Fig.~\ref{Fig5}c, to be a good model for $\theta_P$ and obtain $\theta_S$, 
as plotted in Fig.~\ref{Fig5}d. These results show little variation in 
\textit{S-}wave phase up to about 1.02--1.03~\gevcc, then a rapid rise above 
that. Also, in Fig.~\ref{Fig3}b, we observe that $\left| P \right|$ follows 
the $\phi(1020)$ curve well up to about the same mass, with a significant 
deviation above that.  The behavior observed matches well to that obtained 
from the isobar model I or II. No distinction between them appears possible 
from this analysis. The partial-wave analysis described above is valid, in 
the absence of higher spin states, only if no interference 
occurs from the crossing $K\pi$ channels. The behavior observed in both 
\textit{S-} and \textit{P-}waves above $\sim$1.03~\gevcc\ can, therefore, be 
attributed to high mass tails of the $K^*(892)$ and low mass tails of 
possible higher $K^*$ resonances.\\
\begin{figure}[!htbp]
\includegraphics[width=0.5\textwidth]{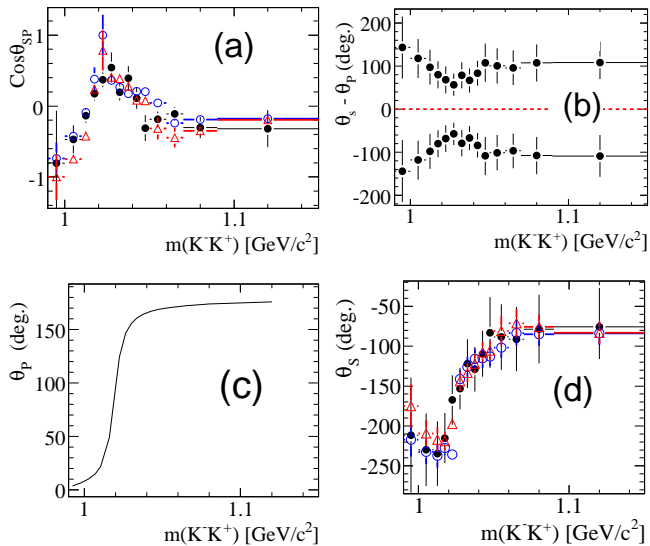}
\caption{(Color online) Results of the partial-wave analysis of the $K^-K^+$ 
system using Eq.~\ref{eq:8} described in the text. (a) Cosine of relative
phase $\theta_{SP}=\theta_\textit{S} - \theta_P$, (b) two solutions for 
$\theta_{SP}$, (c) \textit{P-}wave phase taken from 
Eqs.~\ref{eq:3}--\ref{eq:4} for the $\phi(1020)$ meson, and (d) 
\textit{S-}wave phase derived from the upper solution in (b). Solid bullets 
are data points, and open circles (blue) and open triangles (red) correspond, 
respectively, to isobar models I and II. The number of simulated events used 
for the two models is 10 times larger than data. Errors for 
quantities from the isobar models arise from Monte Carlo statistical 
limitations, and differ from errors derived from Eq.~\ref{eq:8}.}
\label{Fig5}
\end{figure}
\indent Systematic uncertainties in quantities in Table~\ref{tab:result} 
arise from experimental effects, and also from uncertainties in the nature
of the models used to describe the data. We determine these   
separately and add them in quadrature. In both cases, we assign the 
maximum deviation in the observed quantities (i.e., $a_r$, $\phi_r$, and  
$f_r$) from the central value as a systematic uncertainty, taking 
correlations among fit parameters into account. We characterize the 
uncertainties due to $K\pi$ \textit{S-}wave amplitudes and resonance 
mass-width values as model dependent. We estimate them conservatively taking 
symmetric errors from the spread in results when either the LASS amplitude is 
replaced by the E-791 amplitude, or the resonance parameters are changed by 
one standard deviation ($\sigma$). Similarly, we estimate the experimental 
uncertainty from the variation in results when either the signal 
efficiency parameters are varied by 1$\sigma$, or the background shape is 
taken from simulation instead of the data sideband, or  
the ratio of particle-identification rates in data and 
simulation is varied by 1$\sigma$. Model and experimental systematics 
contribute almost equally to the total uncertainty. 
As a consistency check, 
we analyze disjoint data samples, in bins of reconstructed \Dz\ mass and 
laboratory momentum, and find consistent results.\\
\indent Neglecting $C\!P$ violation, the strong phase difference, $\delta_D$, 
between the \Dzb and \Dz decays to $K^*(892)^{+}K^-$ state and their amplitude 
ratio, $r_D$, are given by
\begin{equation}
r_D e^{i\delta_D} = \frac{a_{\Dz\to K^{*-}K^+}}{a_{\Dz\to K^{*+}K^-}} 
{ } e^{i(\delta_{K^{*-}K^+}{ } - { } \delta_{K^{*+}K^-})}. 
\end{equation}
Combining the results of models I and II, we find $\delta_D$ = 
$-35.5^\circ \pm 1.9^\circ$ (stat) $\pm 2.2^\circ$ (syst) and $r_D$ = 0.599 
$\pm$ 0.013 (stat) $\pm$ 0.011 (syst). These results are consistent with the 
previous measurements~\cite{cleo}, $\delta_D$ = $-28^\circ\pm 8^\circ$ (stat) 
$\pm 11^\circ$ (syst) and $r_D$ = 0.52 $\pm$ 0.05 (stat) $\pm$ 0.04 (syst).\\
\indent In conclusion, we have studied the amplitude structure of the decay 
\Dzkkpz, and measured $\delta_D$ and $r_D$. We find that two isobar models 
give excellent descriptions of the data. Both models include 
significant contributions from $K^*(892)$, and each indicates that  
$\Dz\to K^{*+}K^-$ dominates over $\Dz\to K^{*-}K^+$. This suggests that, in 
tree-level diagrams, the form factor for \Dz\ coupling to $K^{*-}$ is 
suppressed compared to the corresponding $K^-$ coupling. While the measured 
fit fraction for $\Dz\to K^{*+}K^-$ agrees well with a phenomenological 
prediction~\cite{theory} based on a large SU(3) symmetry breaking, the 
corresponding results for $\Dz\to K^{*-}K^+$ and the color-suppressed 
$\Dz\to\phi\pi^0$ decays differ significantly from the predicted values.
It appears from Table~\ref{tab:result}
that the $K^+\pi^0$ \textit{S-}wave amplitude can absorb any $K^*(1410)$ and   
$f_2'(1525)$ if those are not in the model. The other components are quite 
well established, independent of the model.
The $K\pi$ \textit{S-}wave amplitude is consistent with that from 
the LASS analysis, throughout the available mass range. We cannot, however, 
completely exclude the behavior at masses below $\sim$1.15~\gevcc observed in 
the decay $D^+\to K^-\pi^+\pi^+$~\cite{kappa, brian}. The $K^-K^+$ 
\textit{S-}wave amplitude, parametrized as either $f_0(980)$ or $a_0(980)^0$, 
is required in both isobar models. No higher mass $f_0$ states are found to 
contribute significantly.
In a limited mass range, from threshold up to 1.02~\gevcc, we measure this 
amplitude using a model-independent partial-wave analysis. Agreement with 
similar measurements from $\Dz\to K^-K^+\bar{K^0}$ decay~\cite{antimo}, and 
with the isobar models considered here, is excellent.\\ 
\indent We are grateful for the excellent luminosity and machine conditions
provided by our PEP-II colleagues, and for the substantial dedicated effort 
from the computing organizations that support \babar. The collaborating 
institutions wish to thank SLAC for its support and kind hospitality. This 
work is supported by DOE and NSF (USA), NSERC (Canada), CEA and
CNRS-IN2P3 (France), BMBF and DFG (Germany), INFN (Italy), FOM (The 
Netherlands), NFR (Norway), MIST (Russia), MEC (Spain), and PPARC (United 
Kingdom). Individuals have received support from the Marie Curie EIF (European 
Union) and the A.~P.~Sloan Foundation.


\begin{thebibliography}{99}
\bibitem{note1}
Reference to the charge-conjugate decay is implied throughout. The initial 
state referred to is \Dz, not \Dzb.
\end{thebibliography}
\end{document}